\documentstyle[prl,twocolumn,aps]{revtex}
\begin{document}
\draft
\narrowtext
\noindent
{\large \bf Density-relaxation part of the self energy}

\vskip 3mm

In a recent Letter, \"{O}\u{g}\"{u}t, Chelikowsky and Louie~\cite{OCL}
presented an important series of calculations of the effect of quantum
confinement on optical gaps in large hydrogen-passivated spherical
silicon clusters, by calculating the quasiparticle energies for
addition of an electron and of a hole separately, and then the
excitonic binding energy.
The quasiparticle energies were calculated by what might be termed a
$\Delta$LDA approach: within the local-density approximation (LDA),
the ground-state total energies of the $n$-, $n-1$- and $n+1$-electron
systems (where $n$ is the number of electrons in the neutral cluster)
were calculated, and then the quasiparticle gap was estimated using

\begin{equation}
\epsilon_{\rm g}^{\rm qp} = E_{n+1} + E_{n-1} - 2 E_{n} .
\label{EqEG}
\end{equation}
The authors suggest that this expression would be expected to approach
the experimental quasiparticle energy gap of bulk silicon (1.2 eV) in
the large-cluster limit.  They presented a numerical fit of the
correction 
$\epsilon_{\rm g}^{\rm qp} - \epsilon_{\rm g,LDA}^{\rm qp}$, which
they stated approached the bulk value of 0.68 eV like $d^{-1.5}$,
where $\epsilon_{\rm g,LDA}^{\rm qp}$ is the LDA Kohn-Sham eigenvalue
gap and $d$ is the cluster diameter.
However, it is known that in the bulk limit Eq.~(\ref{EqEG}) (in the
LDA) simply yields the LDA energy gap: the correction is zero.  This
is because the LDA exchange-correlation energy is an analytic
functional of density: the fact that the change in electron density on
adding (or subtracting) a single electron is of order $1/n$ allows the
changes in the Kohn-Sham eigenvalues and the other ingredients of the
energy to be evaluated using perturbation theory, and after a
substantial cancellation between terms the stated result is obtained. 
(The same formula yields the {\it correct} gap in exact Kohn-Sham DFT,
but this reflects a non-analytic discontinuity in the
exchange-correlation potential between the $n$- and $n+1$-electron
systems~\cite{SSPL}.)

\vskip 3mm

In physical terms, the $\Delta$LDA approach includes the electrostatic
effect of the relaxation of the charge density when an electron is
added or subtracted, and the corresponding relaxation in the LDA
exchange-correlation potential.  In the large-cluster limit, both
these effects go to zero, and the non-zero band-gap correction may be
calculated using many-body perturbation theory in a suitable
approximation (e.g.~\cite{HL,GSS}), where the correction to the
LDA band gap arises from the differing effects of the non-local
self-energy on the states concerned~\cite{GSS}.

\vskip 3mm

Furthermore, there is no reason to suppose that this term in the
self-energy correction that is excluded in the $\Delta$LDA approach is
negligible in the clusters studied.  Therefore it is likely that the
quasiparticle and optical gaps given in Ref.~\cite{OCL} should be
increased by very approximately 0.68~eV, where the error bar in this
estimated correction is smallest for the largest clusters.  Of course,
this additional correction is of lower relative importance for the smaller
clusters.

\vskip 3mm

To confirm our theoretical analysis, we have reanalyzed the data for
the $\Delta$LDA gap correction from Ref.~\cite{OCL}, shown here in
Fig. 1 as a function of $1/d$.  The dashed curve shows the best
(least-squares) fit of the form 0.68~eV + $A d^{- p}$, as in
Ref.~\cite{OCL}, obtained by us with $p=$1.40 (similar to the 1.5
given in Ref.~\cite{OCL}).  The solid curve shows the best fit
obtained if the constraint that the limit as $d \rightarrow \infty$
should be 0.68 eV is removed (as it should be): $K+A d^{- p}$ with
$K$=(0.12$\pm$0.24) eV, $p=0.92 \pm 0.14$.  The value of $K$ is indeed
consistent with zero, and inconsistent with 0.68 eV.  The second fit
is more than twice as good as the first, as measured by $\chi ^2$. 

\vskip 3mm
\noindent R.W. Godby$^a$, and I.D. White$^b$

{\small $^a$Department of Physics, University of York, Heslington,
York YO1 5DD, U.K.}

{\small $^b$Cavendish Laboratory, University of Cambridge, Cambridge
CB3 0HE, U.K.}

\section*{Figure Captions}

\begin{enumerate}

\item[Fig.1]

The band-gap correction from Ref.~\cite{OCL}, plotted against the
inverse cluster diameter.  The best fit (solid curve) correctly tends
to a value consistent with zero.

\end{enumerate}

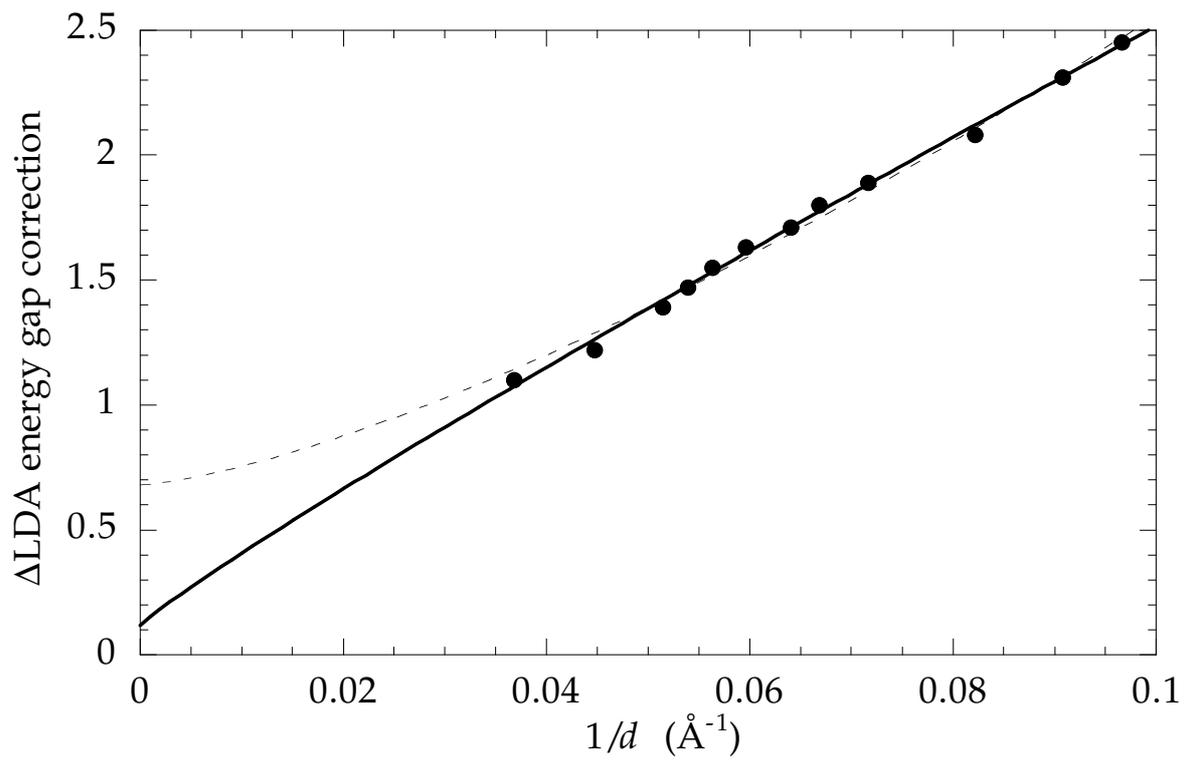
\end{document}